\documentclass[trackchanges]{aastex6}
\usepackage{natbib}
\usepackage{graphicx}
\usepackage{color}
\usepackage{amsmath}
\usepackage{epstopdf}
\begin{document}
	
\title{A possible mechanism for driving oscillations in hot giant planets}
\author{Ethan Dederick}
\affil{New Mexico State University}
\email{dederiej@nmsu.edu}

\author{Jason Jackiewicz}
\affil{New Mexico State University}
\email{jasonj@nmsu.edu}

	\maketitle
	
	\section{Abstract}
	The $\kappa$-mechanism has been successful in explaining the origin of observed oscillations of many types of ``classical'' pulsating variable stars. Here we examine quantitatively if that same process is prominent enough  to excite the potential global oscillations within Jupiter, whose energy flux is powered by gravitational collapse rather than nuclear fusion. Additionally, we examine whether external radiative forcing, i.e. starlight, could be a driver for global oscillations in  hot Jupiters orbiting various main-sequence stars at defined orbital semimajor axes. Using planetary models generated by the Modules for Experiments in Stellar Astrophysics (MESA) and nonadiabatic oscillation calculations, we confirm that Jovian oscillations cannot be driven via the $\kappa$-mechanism. However, we do show that in hot Jupiters oscillations can likely be excited via the suppression of radiative cooling due to external radiation given a large enough stellar flux and the absence of a significant oscillatory damping zone within the planet. This trend seems to not be dependent on the planetary mass. In future observations we can thus expect that such planets may be pulsating, thereby giving greater insight into the internal structure  of these bodies.
	
	\section{Introduction}
	Since the advent of helioseismology and the detection of global solar modes,  it has been postulated that gas-giant planets could also exhibit similar oscillations.  Theoretical work has been done investigating the inherent nature of these oscillatory modes in the giant planets \citep[e.g.,][]{Vorontsov1976,Marley1991,Mosser1995,Jackiewicz2012}, and indeed, they may even have been recently observed. \citet{Gaulme2011} demonstrated a preliminary detection of Jupiter's global acoustic oscillations, and \citet{Hedman2013}  have measured spiral density structures in Saturn's rings caused by Saturnian surface-gravity waves. However, the driving mechanism of these oscillations remains a challenge to understand. \citet{Goldreich1977} demonstrated that the energy from turbulent convection could drive the observed solar oscillations, yet a similar approach for Jupiter and Saturn reveals that too little energy is available from turbulent convection to be responsible for driving their global oscillations \citep{Gaulme2014}. The ratio of velocity of the convective flux to the sound speed (Mach number) gives an indication as to the energy in a driving mechanism. In Jupiter, the Mach number relative to the Sun is much lower, resulting in oscillatory surface amplitudes of at least three orders of magnitude less than the Sun \citep{Bercovici1987}. Therefore, a different source mechanism must be at least partially responsible for exciting oscillations to detectable amplitudes.

	 Several other possible sources can be considered that could potentially drive Jovian global oscillations. These include moist convection in the upper atmosphere, ortho- to para-hydrogen conversion, and helium rain \citep{Bercovici1987}. Yet another source could be the well-understood $\kappa$-mechanism, at work in certain classes of pulsating variable stars \citep{Cowling1941,Stothers1976}. Some of the pioneering work done to develop this theory was conducted by \citet{Cowling1941} and \citet{Eddington1941}, the latter resolving more nuanced questions such as hydrogen ionization being responsible for the phase-retardation in the pulsations.  While the effectiveness of the $\kappa$-mechanism typically requires (partial) ionization regions (e.g., hydrogen and helium) within a star to successfully drive oscillations, we explore if  in the Jovian case, this mechanism could operate at some level. The ongoing contraction of Jupiter releases non-negligible internal radiation (radiating towards the surface) which could be absorbed by any opacity features in Jupiter's atmosphere, thereby having a similar effect as hydrogen or helium ionization zones in pulsating variable stars. 

         We study this possibility using interior models of Jupiter and nonadiabatic pulsation analysis. We then apply the same strategy to look at gas-giant planetary models that are orbiting close to a host star, so-called hot Jupiters. The goal is to understand if the high levels of irradiation impact any potential pulsation driving. 	In Section~\ref{kap} we review the physical description of the excitation of nonadiabatic pulsations in the context of the $\kappa$ mechanism.  Section~\ref{mod} explains the process by which our planetary models are created and calibrated. Section~\ref{analysis} details our analysis of  Jovian and hot Jupiter oscillations, and we end with concluding remarks in Section~\ref{conclude}.


\section{The Kappa Mechanism}\label{kap}
Within a star (or planet), hydrostatic disequilibrium is an imbalance between the gravitational force radially inward and the hydrostatic pressure radially outward. This is an unstable state, causing the star or planet to attempt to re-establish hydrostatic equilibrium.  Consider a small perturbation  in the object in some mass shell.  Assume that an imbalance between pressure and gravity causes the star to collapse a small amount. If this collapse occurs in one or more ionization regions of the star, the rise in density and temperature results in an  increase in the radiative opacity.

Once maximum compression has been reached, the temperature of the compressed region remains constant, yet the increased opacity causes increased photonic heating. The extra heat causes an increase in entropy which forces the gas to now expand isothermally. It is this heating by photons that continues to increase the pressure of the gas while the density now begins to decrease. This results in a phase retardation between the pressure and density. As the expansion proceeds, the effects of photonic heating become negligible and the gas resumes expanding adiabatically. Due to the excess entropy, however, once expansion has finished, the formerly compressed gas now has a larger volume than when it originally began compressing.

At maximum expansion, the gas cools to the surrounding temperature and recompresses isothermally to its original volume. Compressive overshoot occurs and the process likely repeats. For a closed cycle, the entropy must return to its original value. In order for this to be possible, either the gas must spontaneously heat up (which does not occur) or the excess heat must be converted to work. It is this work that is then available to drive  oscillations in a  star.

Thus, the emergent luminosity that normally would  escape the layer easily, because of the fact that  a temperature and density increase upon compression would cause a reduction in the opacity,  instead results in partially ionizing  the gas, increasing opacity, and  contributing a net positive energy into the cycle at maximum compression. This is the principle driving behind the $\kappa$-mechanism.


To quantify the amount of work  available to drive oscillations, the work function is the mathematical expression that describes the energy transfer between oscillations and the medium.  There are many excellent derivations of the work integral approach  \citep{Cox1980,Unno1989,Pamyatnykh1999,Cunha2001,Samadi2015}, to which we refer the reader for details. For our purposes, we consider the expression \citep[as in][]{Samadi2015}
	 \begin{equation}
	 	 \Bigg<\frac{dW}{dt}\Bigg>  = \Re\Bigg[\int\limits_{0}^{M}\Big(\Gamma_3 -1\Big)\Bigg(\frac{\delta\rho}{\rho}\Bigg)^*\Bigg(\delta\epsilon - \frac{d}{dm}\delta L\Bigg) \Bigg]dm.
	 \label{eq:workfunc}
	 \end{equation}
 	 Here $\Re$ indicates the real part of a complex number and $*$ represents its complex conjugate. $\Gamma_3$ is one of the adiabatic exponents relating temperature and density, $\rho$ is density, $\epsilon$ is the rate of energy generation (negligible for planets), and $L$ is luminosity. Lagrangian perturbations to these quantities due to oscillations are denoted with a $\delta$.

Equation~(\ref{eq:workfunc})  describes the power supplied to or subtracted from an oscillation averaged over one pulsation cycle by the surrounding gas. If the total work available is positive, this energy can be transferred non-adiabatically to driving (unstable) oscillations. Otherwise, the oscillations are stable and damped. For our purposes, we will be exploring the conditions necessary for a positive work integral, ignoring perturbations to the energy generation rate, which are negligible. In particular, for the modes of interest, we will find that $\Gamma_3-1$ is always greater than zero and that density perturbations are negative. Therefore, positive luminosity perturbations induced by a disturbance can provide an overall positive work function. If an opacity effect is responsible for the positive perturbation, it is the $\kappa$-mechanism.


	 
	 \section{Models}\label{mod}
	 \subsection{Calibration}

To study the stability of pulsations, we developed a code that computes the work function in Eq.~(\ref{eq:workfunc}) from the  profiles of 1D stellar and planetary models and their associated perturbations. For the 1D models, we employ the MESA (Modules for Experiments in Stellar Astrophysics) stellar evolution code \citep{Paxton2011,Paxton2013,Paxton2015}. For our purposes, MESA uses the calculations of \citet{ferguson2005} to implement opacities in the low-temperature regimes necessary for gas giants. These models  are then used as input for the GYRE suite of stellar oscillation codes \citep{Townsend2013}, which utilizes a multiple shooting scheme to solve the non-adiabatic oscillation differential equations to produce a spectrum of eigenfrequencies, eigenfunctions, and perturbed quantities.

 These tools were first tested by replicating the differential work functions of models for two classes of well-studied stars that exhibit pulsations via the $\kappa$-mechanism:  $\beta$ Cepheid stars and  $\delta$ Scuti stars. More specifically, we calibrate by trying to match the work function in Figure 2 of \citet{Pamyatnykh1999}, particularly the $l=1$, p1 (n=1) and p4 (n=4) modes for the $\beta$ Cep star and the $l=1$, p1 and p7 (n=7) modes for the $\delta$ Sct star. $l$ is the angular degree and $n$ is the radial order of the acoustic oscillations. 

\begin{figure}[t]
	 	\centering
                \centerline{
	 	\includegraphics[width=.5\textwidth]{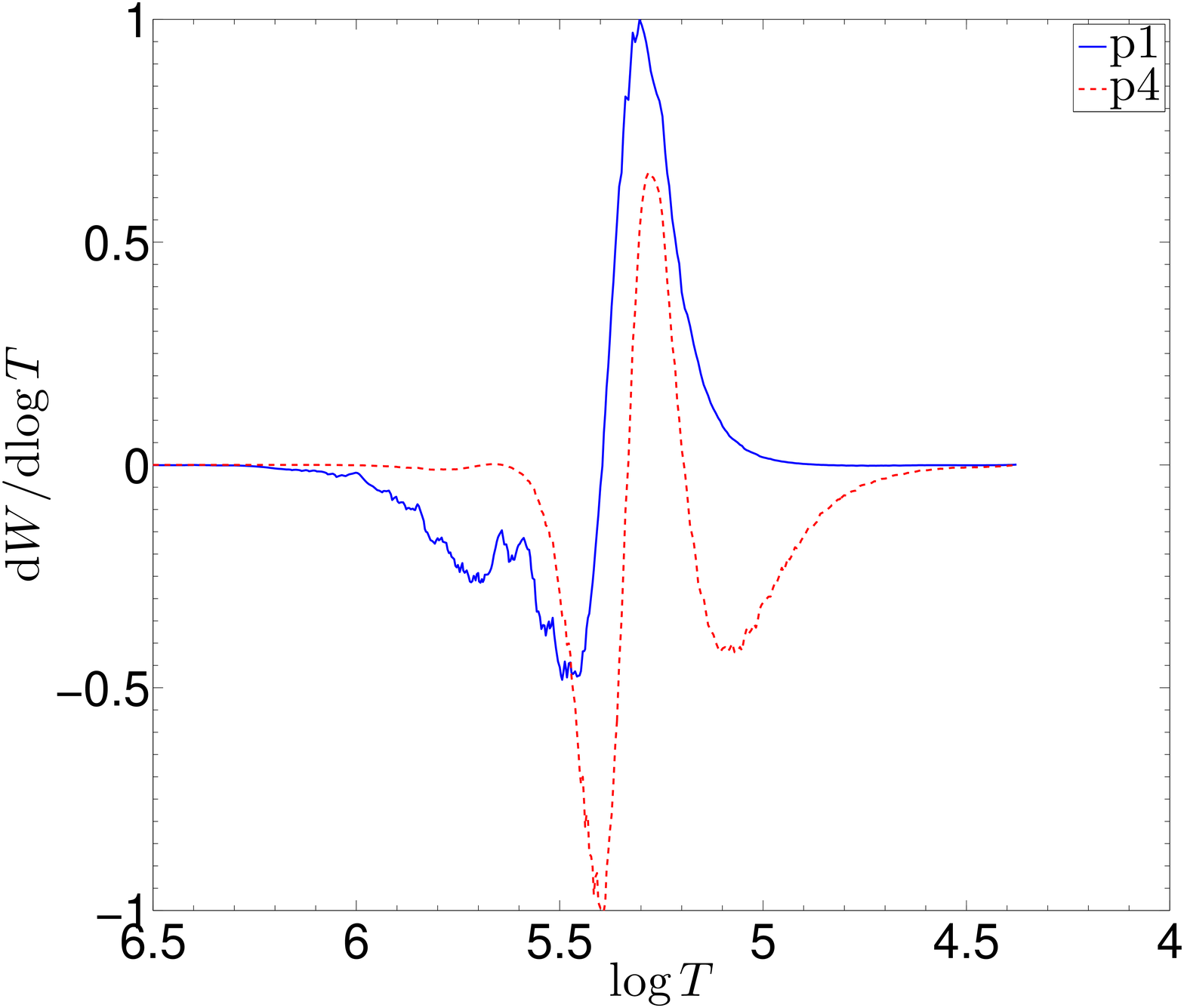}
                \includegraphics[width=.5\textwidth]{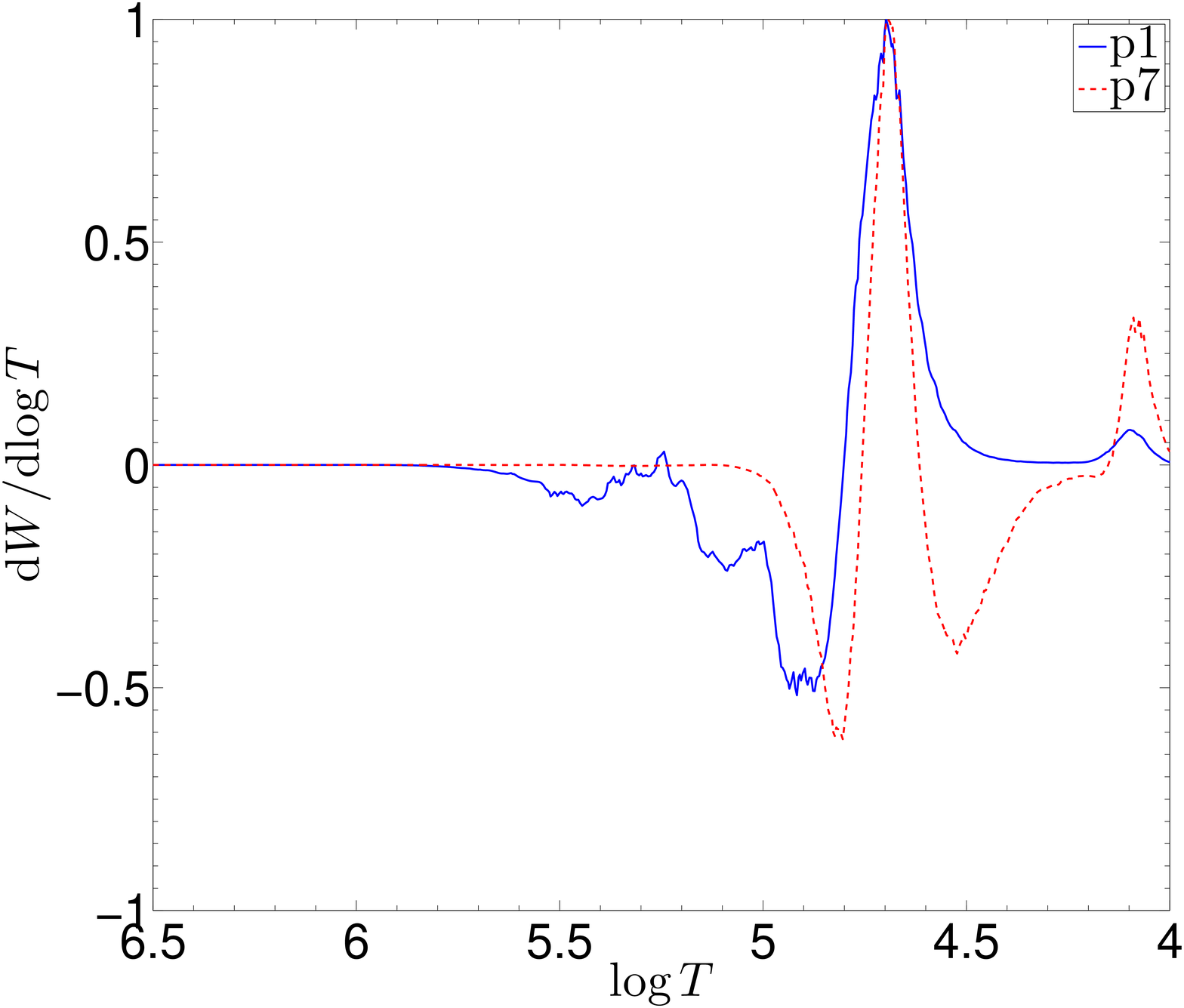}}
                \caption{Left: The differential work function for a $\beta$ Cep model and the  $l=1$, p1 and p4 modes. Work function units are arbitrary. p1 is an unstable mode while p4 is stable.
	 Right: The differential work function for a $\delta$ Sct model and the $l=1$, p1 and p7 modes.  p1 is an unstable mode while p7 is stable. Temperatures are in Kelvin.}
	 	\label{fig:delta}
	 \end{figure}

	 Each stellar model is created using the parameters listed in \citet{Pamyatnykh1999}, and the frequencies of the modes calculated with MESA and GYRE are within a few percent of those published values ($\beta$ Cep: 61.55 $\mu$Hz \& 106.18 $\mu$Hz for p1 and p4, respectively; $\delta$ Scu: 170.66 $\mu$Hz \& 384.28 $\mu$Hz for p1 and p7, respectively).  The plots of the differential work function given by Eq.~(\ref{eq:workfunc}) are shown in  Fig.~\ref{fig:delta} for each star and  are nearly identical to those found in \citet{Pamyatnykh1999}.
	 
Oscillations are excited (instability) when the work function is greater than zero. Thus, in Fig.~\ref{fig:delta}, if the area under the curve of the differential work is positive, that particular mode is excited, otherwise it is damped. Positive peaks correspond to driving regions, while negative peaks are damping regions. Therefore, in the $\beta$ Cep star, the p1 (p4) mode is unstable (stable), and likewise, in the $\delta$ Sct star, the p1 (p7) mode is unstable (stable), which matches the results in \citet{Pamyatnykh1999}. It must be noted that a positive work function  does not necessarily mean that a mode is driven. It may be excited, but it must also have a positive growth rate or else it will be quickly damped out. A more detailed discussion of mode growth rates is given in Section~\ref{growth}. Finally, the driving regions must also be near the surface, as perturbations deep in the stellar interior must be exceedingly large to overcome the thermal inertia of the surrounding medium. 
	 
	 \subsection{Jupiter Models}
To study the behavior of any non-adiabatic oscillations excited by the $\kappa$-mechanism within Jupiter, we use MESA following this prescription:
\begin{enumerate}
\item
  A gas giant planet is created with a radius of 1.2 - 2 Jupiter radii with mass fractions $X=0.74$, $Y=0.24$, $Z=0.02$. It has a mass of 1 Jupiter mass minus the mass of a currently non-existent core. The model is then evolved for 1000 years to allow for relaxation of the internal equation of state.
\item
  A core with an average density of ${\rm 10\,g\,cm^{-3}}$ and the specified mass needed to bring the planet total to one Jupiter mass, is slowly added over 2000 years to allow the equation of state to relax once more.
\item
  The model is evolved over 4.5 Gyr, allowing the interplay between gravity and internal pressure to evolve the equation of state. During this time, the surface of the planet may be irradiated with some specified bolometric solar flux, penetrating to a specified maximum column depth. At the end of the evolution, the radius of the planet has slightly decreased due to gravitational collapse.
\item
  Each model is then fed into GYRE where it is linearly scanned for oscillations between 0.1 and 5 mHz for the $l=0,1,2$ angular degrees and radial orders $n=1-10$. The output from both GYRE and MESA includes all the necessary components of Eq.~(\ref{eq:workfunc}). It can then be determined which modes, if any, are potentially excited within Jupiter.
\end{enumerate}

\begin{table}[t]
  \centering
  \caption{Parameters of the Jovian models.}
  \begin{tabular}{|c|c|c|c|}
    \hline
    Model Name & Irradiation & Core Mass ($M_\Earth$) & Radius ($R_J$) \\ \hline
    A1         & High       & 0                    & 1.1045      \\ \hline
    A2         & Solar      & 0                    & 1.1045      \\ \hline
    B1         & High       & 10                   & 0.9691      \\ \hline
    B2         & Solar      & 10                   & 0.9691      \\ \hline
    C1         & High       & 15                   & 0.9532      \\ \hline
    C2         & Solar      & 15                   & 0.9532      \\ \hline
  \end{tabular}
  \label{tab:JupMods}
\end{table}
	 
Table~\ref{tab:JupMods} lists six particular Jupiter models that are explored. It lists whether the planet is irradiated with $\sim 10^9$ erg cm$^{-2}$ s$^{-1}$ (High) or $5.03\times 10^4$ erg cm$^{-2}$ s$^{-1}$ (Solar) of irradiation, its core mass, and its final radius after 4.5 Gyr. Each model is irradiated to a column depth of 300 cm$^2$ g$^{-1}$ ($\approx 0.9967$ R$_J$). How far the flux penetrates into the Jovian atmosphere is unknown, hence the column depth of irradiation is somewhat arbitrary. We ran an additional 810 models to investigate the dependence of mode excitation on the column depth of irradiation. By varying the column depth throughout these models, we found that even for various flux values (greater than $10^9$ erg cm$^{-2}$ s$^{-1}$), oscillations can be excited as long as the column depth exceeds 50 cm$^{-2}$ g$^{-1}$.  For the set of models in Table~\ref{tab:JupMods}, 300 cm$^2$ g$^{-1}$ corresponds to a pressure depth of $\sim 1.5$ bars (the surface is defined at $\sim 0.3$ bars). \citet{Gierasch2000} cite observations to a depth of $\sim 3$ bars of pressure for infrared observations. If IR light can be observed radiating from the 3 bar level we assume it can penetrate to the 3 bar level (or a column depth greater than 600 cm$^2$ g$^{-1}$). Since shorter wavelength light will be more readily absorbed, we split the difference between our lower and upper limits and select 300 cm$^2$ g$^{-1}$. Therefore, our estimate of $\sim 1.5$ bars may be somewhat conservative. Finally, it is important to note that these are one-dimensional models and thus assume spherical symmetry. Therefore, we must also assume these are rapidly rotating (i.e. on the order of the rotation periods of Jupiter and Saturn) planets such that any consequences of stellar radiation only penetrating half of the planet's surface at a given time can be ignored. However, this may not be a large concern as a tidally locked planet with the driving region always facing the host star could still experience global oscillations just as how Jupiter oscillates due to (probable) excitations of localized moist convective storms.
	
The high amount of irradiation applied to models A1, B1, and C1 is about five orders of magnitude larger than the solar flux at Jupiter.  We had initially made a rather trivial mistake implementing an incorrect amount of flux in the models and subsequently found surprising results (explained in Section~\ref{analysis}). We thus decided to keep these 3 models for the analysis (even though they don't represent Jupiter proper), and furthermore, were motivated to consider the  more realistic cases of  hot Jupiters orbiting very close to their host star.


\subsection{Exoplanetary Hot Jupiter Models}\label{exo}

We  generated a set of  exoplanet hot Jupiter models that are created using the same method as described above, with a few differences. Table~\ref{tab:stars} lists the subset of main-sequence stellar models selected as host stars for these planets, their masses, their effective temperatures, their bolometric luminosities and their ages. The stellar ages are determined by running MESA main-sequence models until $10\%$ of their core hydrogen abundance is depleted, or 3 Gyr, whichever comes first. The planetary models are then evolved to the same age as the corresponding star.	 

Representative planetary mass and orbital semimajor axis ranges are taken from NASA's exoplanet archive.\footnote{\url{http://exoplanetarchive.ipac.caltech.edu/}} We created a parameter space of masses ranging from 1 to 30 Jupiter masses and orbital semimajor axes of 0.01 to 2 AU. The irradiation flux received by each planet is then the bolometric flux at the particular semimajor axis calculated from a stellar blackbody given by the host star's effective temperature. Each planet model is irradiated to a column depth of 300 cm$^2$ g$^{-1}$ over its entire lifetime.  The models have a core mass of 10 Earth masses with average core density of ${\rm 10\,g\,cm^{-3}}$. We compute planetary models for 9 different host star spectral types with 11 different planet masses at 12 different semimajor axes for a total of 1,188 models. We find that varying the column depth penetration of radiation has negligible effect on the results presented in subsequent sections, as long as it is greater than 100~cm$^2$ g$^{-1}$.

\begin{table}
  \centering
  \caption{Parameters of the stellar models that are host stars to hot Jupiters.}
    \begin{tabular}{|c|c|c|c|c|}
    \hline
    Spectral Type & Mass ($M_\odot$) & $\log T_{\rm eff}$ (K) & Luminosity ($L_\odot$) & Age (Gyr) \\ \hline
    M5            & 0.28            & 3.565         & 0.0126                & 3         \\ \hline
    M0            & 0.49            & 3.584         & 0.0383                & 3         \\ \hline
    K5            & 0.58            & 3.598         & 0.0645                & 3         \\ \hline
    K0            & 0.76            & 3.670         & 0.216                 & 3         \\ \hline
    G5            & 0.91            & 3.731         & 0.523                 & 2.258     \\ \hline
    G0            & 1.05            & 3.767         & 1.027                 & 1.264     \\ \hline
    F5            & 1.33            & 3.816         & 3.144                 & 0.535     \\ \hline
    F0            & 1.59            & 3.871         & 6.828                 & 0.370     \\ \hline
    A5            & 1.98            & 3.944         & 16.700                & 0.203     \\ \hline
  \end{tabular}
  \label{tab:stars}
\end{table}

\section{Analysis}\label{analysis}

	 
	 \subsection{Jupiter oscillations}


We carry out a nonadiabatic oscillation analysis for the Jupiter  models. As one might expect \citep[e.g.,][]{Bercovici1987}, there is not as large a source of internal energy in Jovian planets as there is in stars, and thus we do not find that  the $\kappa$-mechanism (or any other mechanism considered here) excites modes within the nominal Jupiter models. All calculated modes in models A2, B2, and C2 are stable and are thus not excited. Frequencies range between 0.129 mHz and 1.74 mHz depending on the $l$ and $n$ values. 
The differential work function of the models is virtually zero throughout Jupiter except for the outermost $\sim~1\%$ of its radius. In the outermost  layers, there is a strong negative feature, that is, a strong damping region, that contributes to energy lost over an oscillatory cycle. Therefore, these modes are not excited. 
	 
	
	 The introduction of the significant ($\sim 10^9$ erg cm$^{-2}$ s$^{-1}$) solar irradiation throughout Jupiter's evolution, however, results in excited oscillations. In models A1, B1 \& C1, several of the low radial order modes are found to be unstable. For example, model A1 (no core) revealed that  modes $l=0$, $n=1$, \& $l=2$, $n=1$ are excited (recall that angular degrees greater than 2 and radial orders greater than 10 are untested).   The reason the oscillations are excited in the presence of extreme stellar radiation is due to a large luminosity perturbation that occurs at about $R=0.993\ R_J$, a direct consquence of the strong irradiation. A more detailed discussion as to what is happening is given in the following sections for the case of the hot Jupiter models.

	 \subsection{Hot Jupiter oscillations}

\begin{figure}
  \centering
  \includegraphics[width=.6\textwidth]{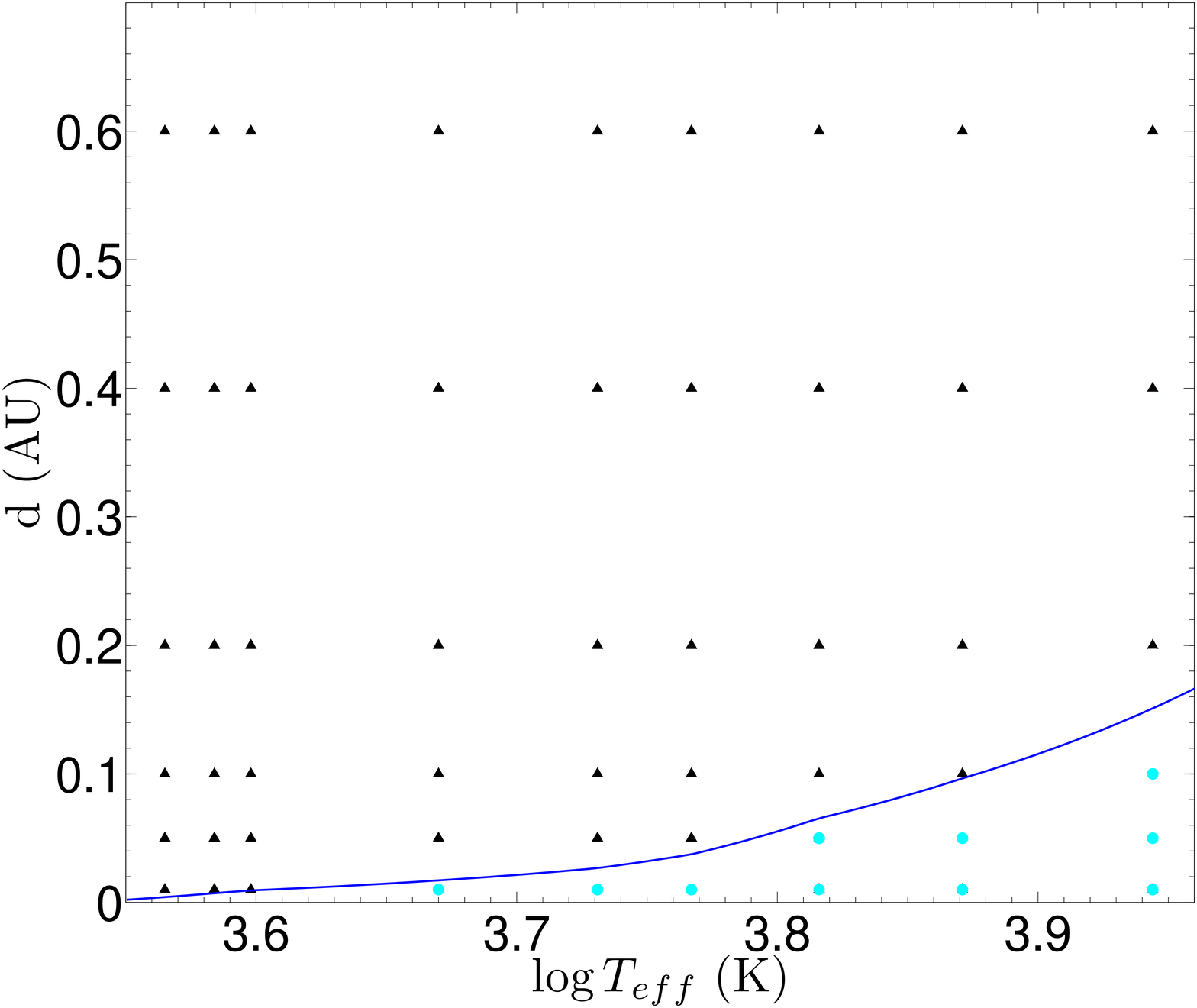}
    \caption{Host-star spectral type (effective temperature) and hot Jupiter orbital distance parameter space.  The small cyan circles represent models with mode instability (excitation) and the large black triangles represent models with mode stability (no excitation). The blue line corresponds to Eq.~(\ref{eq:relation}).  Models were not run at 0.3 or 0.5 AU.}
  \label{fig:HR}
\end{figure}



The highly-irradiated Jupiter models that demonstrate excited modes motivated us to explore  gas giant planets orbiting very close to their host stars. The parameter space described in Section~\ref{exo} reveals some interesting trends regarding which planets exhibit mode excitation.  Figure~\ref{fig:HR} shows the models for which unstable modes are found in terms of orbital distance and host-star effective temperature. We observe a very  well-defined region of excitation favoring  short-period planets or hot host stars.


Regardless of planetary mass, we find that approximately $\sim 10^9$~erg\,cm$^{-2}$\,s$^{-1}$ of flux are required for oscillations to occur. That flux or higher does not guarantee the existence of oscillations, but it is a requirement for them to be present. Thus, the separation line in Fig.~\ref{fig:HR} is given by
\begin{equation}\label{eq:relation}
  d = \sqrt{\frac{\sigma R_*^2T_{\rm eff}^4  }{F}},
\end{equation}
where $d$ is the orbital distance, $\sigma$ is the Stefan-Boltzmann constant, $R_*$ is the stellar radius,  and the flux $F=10^9$~erg\,cm$^{-2}$\,s$^{-1}$. We see that this relation does a good job at discriminating between the planets that have excited modes and those that do not.


	 
\begin{figure}
  \centering
  \includegraphics[width=\textwidth]{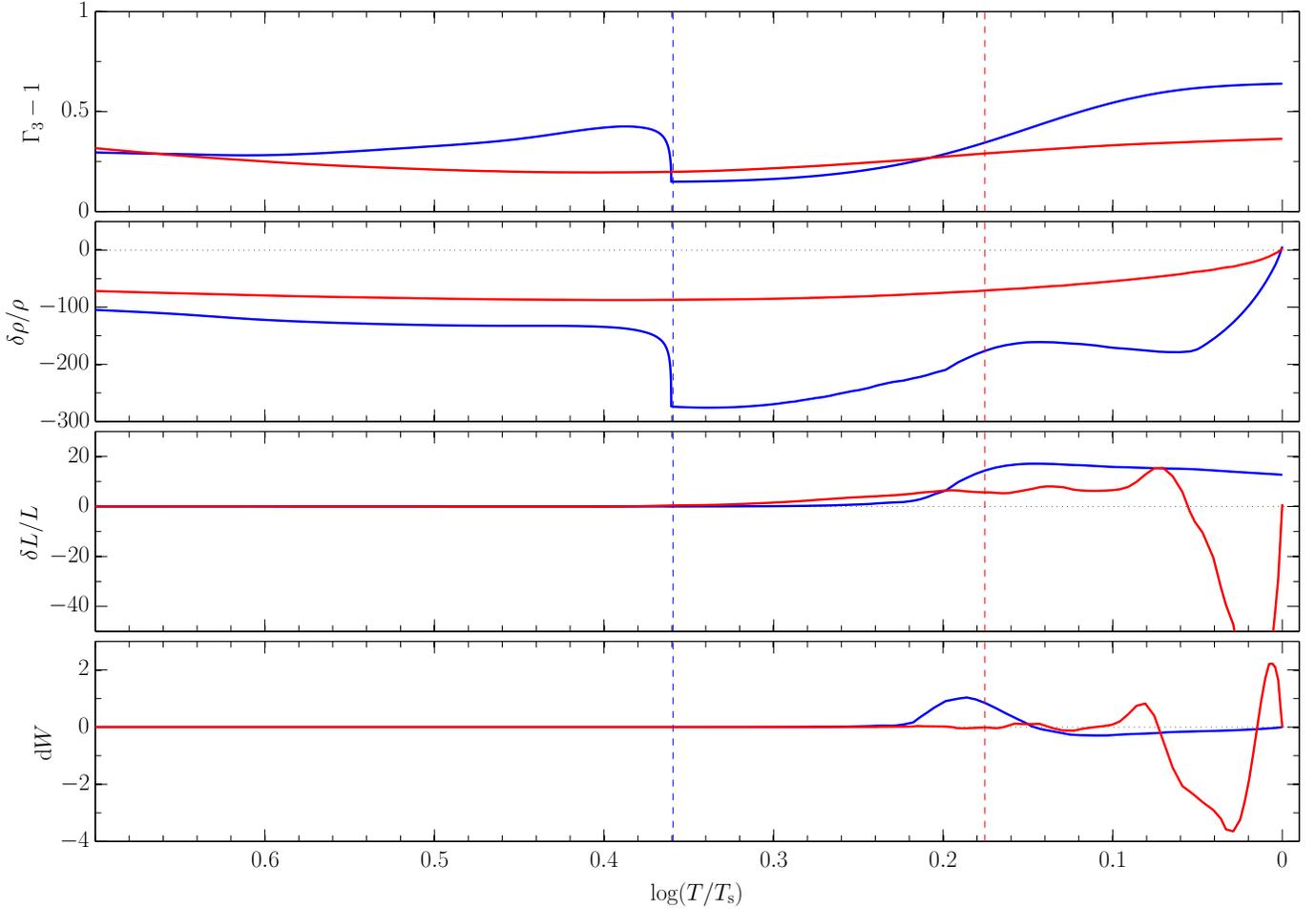}
  \caption{The profiles of the main terms that make up the integrand of the work function in Eq.~(\ref{eq:workfunc}). From top to bottom: adiabatic index, $\Gamma_3 -1$, the density perturbation, the luminosity perturbation, and the differential work function integrand. These correspond to $30\ M_J$ models at distances of 0.01~AU (blue) and 0.2~AU (red) around the A5 star for the $l=0$, $n=2$ mode. The vertical blue (red) dashed line corresponds to the penetration depth of the stellar irradiation for the 0.01~AU (0.2~AU) model. The $x$-axis is the interior temperature profile scaled by the surface value ($T_{\rm s}$), which is about 5600 K and 1600 K for the 0.01~AU and 0.2~AU model, respectively.  The 0.01~AU model planet exhibits excited modes, while the 0.2~AU model planet does not. Vertical axis units in the bottom panel are arbitrary.}
  \label{fig:quad}
\end{figure}

According to Eq.~(\ref{eq:workfunc}), there are three factors that influence the sign of the work function (ignoring perturbations to the energy generation rate).  Figure~\ref{fig:quad} shows the comparison of these three quantities as well as the differential work function between the A5, 30~$M_J$ planets at orbits of 0.01~AU (blue) and 0.2~AU (red), for the $l=0$, $n=2$ mode. This particular mode is excited in the 0.01~AU model, while in the 0.2~AU planet it is not. We choose to display these models in particular as they have the most exaggerated profiles so the effects are easiest to see. 
	 
Throughout the interior $\Gamma_3-1$ is positive and similar in magnitude  between the models and it is not a critical parameter. We see that for this particular solution the density perturbation for this mode is negative in both models (recall that we are solving the time independent equations). So we are at maximum expansion due to the oscillations. To obtain a postive work function, therefore, a large (positive) luminosity perturbation is thus needed, such that the negative gradient of this term is negative. Indeed this is what is found for the close-in planetary model, where it is observed how much further in radius the light penetrates the outer layers of the planet model at 0.01~AU. An oscillation  induces a positive luminosity perturbation over a larger fraction of the planet's radius. 

\begin{figure}
  \centering
  \includegraphics[width=.5\textwidth]{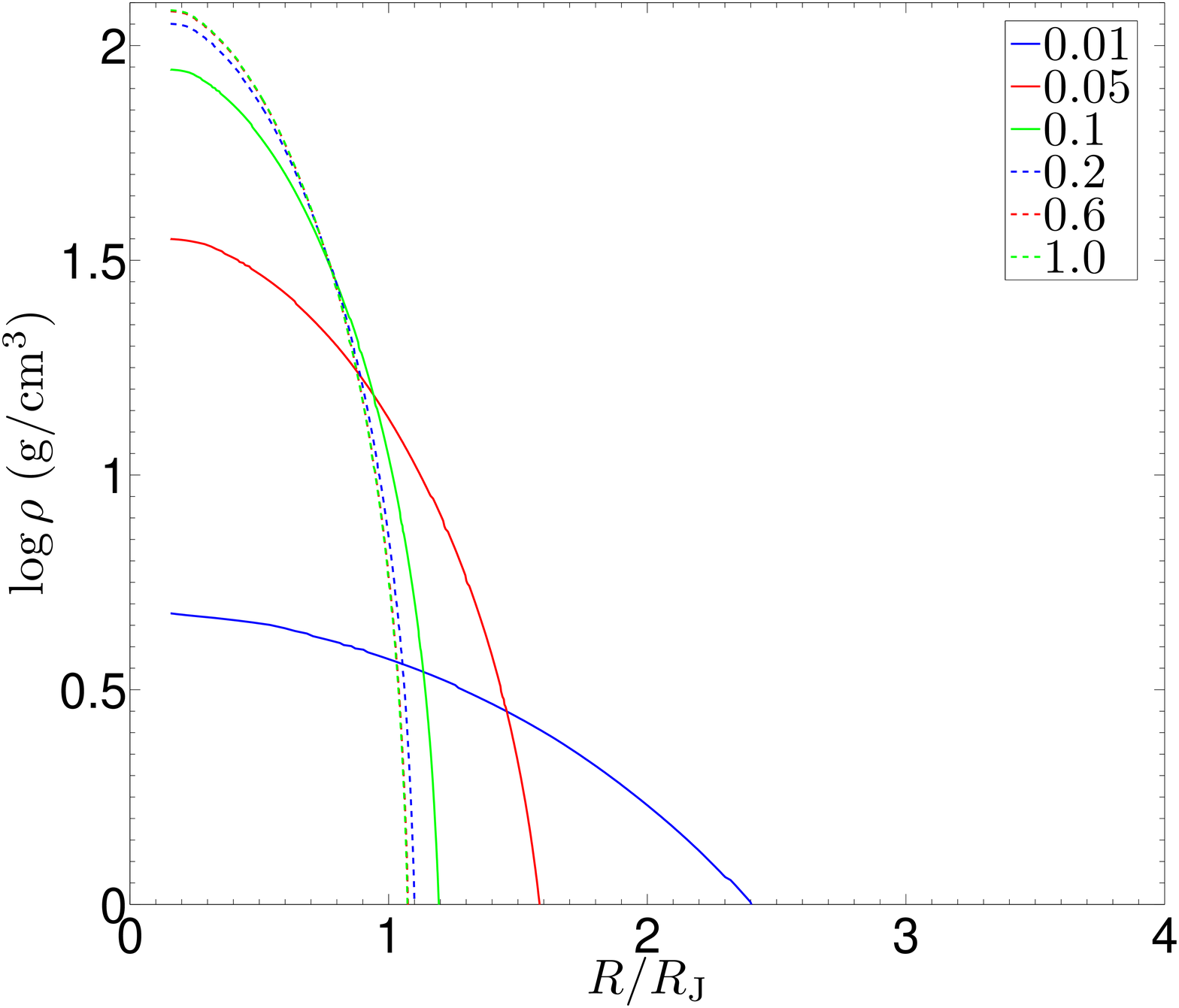}
  \caption{The interior density profiles of the $30\ M_J$ models around an A5 star. The legend indicates planetary orbital semimajor axes in AU. All models have the same age. Note that the innermost data in radius are missing, as that is the region occupied by the core that MESA and GYRE assume to be steady state, so it is left out of the evolved data.}
  \label{fig:rho}
\end{figure}

To explore this situation in more detail, Fig.~\ref{fig:rho} shows the interior density profiles of the $30 M_J$ planets around the A5 star at increasing orbital distance.  As the planet evolves at a further distance from the star, it does not experience significant heating, and therefore becomes a denser planet with a smaller radius. However, the closer the planet is to the star, the more puffed up and diffuse the outermost layers become. This allows light to penetrate deeper into the planet, thereby allowing for a larger region over which significant luminosity perturbations due to modes can occur. Some hot Jupiters have indeed been observed to have larger radii than initially predicted \citep{Bakos2007,Hartman2012}.  The positive luminosity perturbation due to the oscillation heats the gas, causing a negative density perturbation by rarefying it.  This newly expanded gas acts like a restoring force, but it is not large enough to win out over the induced rarefaction by the external heat source. This is a general trend seen over all stars and all mass ranges. 

In addition, for the model with the excited $n=2$ radial mode, its frequency is about $0.3$\,mHz, or a period of an hour. The thermal timescale in the driving region is similar to this, and thus the thermal conditions for the mode are favorable. The driving takes place in a ``transition region''  \citep{Pamyatnykh1999}. This is also the case for all of the modes we consider that are unstable.

Finally, we must also consider the convective timescale in relation to these oscillations. Typically, convection is expected to stabilize the planet against such pulsations. However, the timescale of convection, estimated in a similar fashion as in the case of the Sun (i.e. the dynamical timescale divided by a measure of superadiabaticity), is quite large. In the driving region for the 0.01 AU, 30 Jupiter mass model around the A5 star, for example, the convective timescale is approximately a couple hundred hours, or 2 orders of magnitude larger than thermal timescale and the pulsation period. Therefore, we expect convection to be a passive process regarding these pulsations.

\subsection{Mode Growth Rate}\label{growth}

In addition to  the work function needing to be positive in order to excite modes, as well as a thermal timescale similar to the mode period, we also examined the $\eta$ growth rate parameter as discussed in \citet{Ando1975}. This parameter is given by
\begin{equation}
  \eta = \frac{-\sigma_i}{\sigma_r}
\end{equation}
where $\sigma_i$ is the imaginary component of the (non-adiabatic) mode eigenfrequency and $\sigma_r$ is the real component. If $\eta > 0$ the mode is overstable and will grow, and if $\eta < 0$, the mode is understable and will be damped out. Throughout this analysis, if the work function of a particular mode is positive but $\eta$ is negative, then the mode is considered to be excited but quickly damped. In Fig.~\ref{fig:HR}  the modes that are excited and quickly damped are listed as unexcited data points.

The values for $\eta$, both for the Jupiter and the hot Jupiter modes are mostly within the range calculated by \cite{Ando1975}. Their values range from $10^{-10}$ to $10^{-4}$ whereas our values range from $10^{-14}$ to $10^{-3}$. Yet the best indication as to whether these modes are overstable by examining $\eta$ is actually to compute $\omega/\eta$ where $\omega$ is the frequency of the mode (in mHz). The mode is almost certainly overstable for large values of $\omega/\eta$. We calculate a range of 98.8 mHz to $2\ \times \ 10^{13}$ mHz with the vast majority of the modes at the upper end of that range. 

\subsection{Radiative Suppression}\label{radsuppress}

\begin{figure}
  \centering
  \includegraphics[width=.5\textwidth]{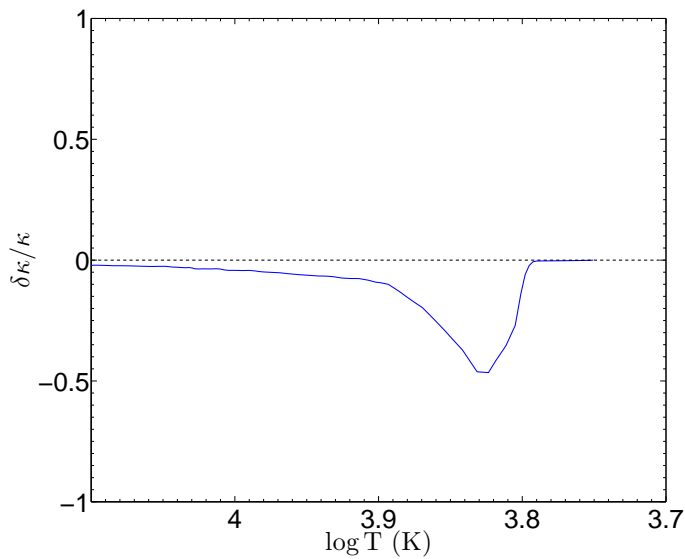}
  \caption{The opacity perturbation due to an unstable $l=0$, $n=2$ mode for the $30\ M_J$ model around an A5 at 0.01~AU. The values have been normalized. The perturbation is at maximum expansion ($\delta\rho<0$).}
  \label{fig:opacperturb}
\end{figure}

The process by which modes are excited via radiative suppression can now be explicitly explained. The combination of the positive (negative) luminosity perturbation and negative (positive) density perturbation is what is leading to unstable modes. The radiative luminosity perturbation can be shown to be \citep{Pamyatnykh1999}:
\begin{equation}\label{eq:lumperturb}
  \frac{\delta L_r}{L_r} = \frac{\textnormal{d}r}{\textnormal{d}\ln T}\frac{\textnormal{d}}{\textnormal{d}r}\Bigg(\frac{\delta T}{T}\Bigg)-\frac{\delta\kappa}{\kappa}+4\frac{\delta T}{T}+4\frac{\delta r}{r}.
\end{equation}
In our case, the first term on the RHS of Eq.~(\ref{eq:lumperturb}) contributes to oscillatory damping and the later two terms are negligible. The dominant term on the RHS is the opacity perturbation, $\delta\kappa/\kappa$,  that drives the oscillations.  For example, when we do detect an unstable mode, we indeed find at maximum expansion that the opacity perturbation is rather large and negative, such that Eq.~(\ref{eq:lumperturb}) is positive. Figure~\ref{fig:opacperturb} shows an example from the same extreme model used earlier.


The mechanism at work is as follows. Some weak perturbation in the gas causes it to collapse at a given location. Upon maximum compression, the density, opacity, and temperature all increase. In normal conditions, the temperature increase allows the gas to cool via radiative losses more readily. However, this cooling is suppressed by the high amounts of external radiation heating the gas due to the increased opacity. Fundamentally, it is this absorption of energy that is converted to work to drive the mode. With the increase in energy from the photonic heating, the gas then begins to expand, pushing past the point of equilibrium and eventually to a state of maximum expansion. 
Here, the opacity perturbation is negative and the gas cools because the luminosity perturbation is positive, increasing outwards (Figs.~\ref{fig:quad} and \ref{fig:opacperturb}). The recompression after maximum expansion overshoots the point of equilibrium and the process repeats.

This process is in contrast to the ``classical'' kappa mechanism in two fundamental ways. First, the kappa mechanism is intrinsically an internally-driven phenomenon with the radiation originating from nuclear fusion whereas the radiative suppression is an externally driven phenomenon. Second, in normal stellar conditions, a compression results in a decrease in opacity rather than an increase. The reason the kappa mechanism works is because in unique regions within the stellar interior, compression raises the temperature of the gas to a level at which hydrogen or helium can be ionized, which thereby increases the opacity. The radiative suppression mechanism does not rely on such ionization as a prerequisite.


\section{Conclusions}\label{conclude}

In this work we developed MESA models of Jupiter and hot Jupiters to ascertain if, and how, nonadiabatic oscillations can be excited. For Jupiter, as expected, oscillations are not excited via the $\kappa$-mechanism, as the intrinsic  luminosity and received solar flux is too small. However, for giant planets orbiting sufficiently nearby  hot stars, a ``radiative suppression'' mechanism can drive global oscillations due to the external stellar irradiation. We find approximately that about $\sim 10^9$ erg cm$^{-2}$ s$^{-1}$ of flux is required, which sets the spectral type of host stars and the orbital distance of the planet.  Modes are excited in planets closer to the host star because the outer atmosphere becomes heated and distended, allowing for light to penetrate deeper into the planet which allows for perturbations to occur over significantly larger regions of the outer layers. This effect does not seem to depend on the total mass of the planet.
	 
Sufficiently large stellar irradiation suppresses a mode's ability to radiatively lose energy, and simultaneously supplies energy to the mode by heating the compressed medium due to an increase in the opacity. This excitation process occurs very near to the surface of the planet: within the outermost 1\% of the radius for the models considered here. Increasing the maximum column depth of irradiation does not significantly alter the results. Modes are still excited in the same location. Further study is needed with more detailed models to understand what makes $~10^9$\,erg\,cm$^{-2}$\,s$^{-1}$ the approximate cutoff stellar flux. Also, a more expansive parameter space is needed with the inclusion of higher order modes to understand the conditions for when  lower order vs. higher order modes are excited.
	 
This study leads us to believe that some hot Jupiters will be pulsating. However, as the mode amplitudes are difficult to estimate, it is unclear what implications this has for the upcoming observational capabilities and detection.


\acknowledgements
The authors would like to acknowledge funding from NASA EPSCoR award \#NNX14AN67A to NMSU,  as well as the New Mexico Space Grant Consortium.

\bibliography{DJ_Mech}

\end{document}